\documentclass[conference]{IEEEtran}
\IEEEoverridecommandlockouts
\usepackage{mathtools}
\usepackage{cite}
\usepackage{amsmath,amssymb,amsfonts}
\usepackage{algorithmic}
\usepackage{graphicx}
\usepackage{textcomp}
\usepackage{xcolor}
\usepackage{algorithm}
\usepackage{algorithmic}
\usepackage[colorlinks,allcolors=black]{hyperref} % Imports a hyperlink to references
\def\BibTeX{{\rm B\kern-.05em{\sc i\kern-.025em b}\kern-.08em
    T\kern-.1667em\lower.7ex\hbox{E}\kern-.125emX}}

\begin{document}

\title{Uplink Power Control in Integrated Access and Backhaul Networks
}

\author{\IEEEauthorblockN{Olalekan Peter Adare\IEEEauthorrefmark{1}, Haitham Babbili\IEEEauthorrefmark{1}, Charitha Madapatha\IEEEauthorrefmark{1}, Behrooz Makki\IEEEauthorrefmark{2}, \IEEEmembership{Senior Member, IEEE}, and \\ Tommy Svensson\IEEEauthorrefmark{1}, \IEEEmembership{Senior Member, IEEE} \\ \IEEEauthorblockA{\IEEEauthorrefmark{1}\textit{Department of Electrical Engineering}, \textit{Chalmers University of Technology}, Gothenburg, Sweden \\ 
\IEEEauthorrefmark{1}\{peterad, haitham\}@student.chalmers.se, \IEEEauthorrefmark{1}\{charitha, tommy.svensson\}@chalmers.se} 
\IEEEauthorblockA{\IEEEauthorrefmark{2}\textit{Ericsson Research}, Gothenburg, Sweden. \IEEEauthorrefmark{2}\{behrooz.makki\}@ericsson.com} }
}

\maketitle
\pagestyle{plain}
\begin{abstract}
Integrated access and backhaul (IAB) network is a novel radio access network (RAN) solution, enabling network densification for 5G and beyond.
In this paper, we use power control combined with resource allocation algorithms to develop efficient IAB networks with high service coverage. 
Particularly, we develop a genetic algorithm-based solution for the power control of both user equipments and IAB nodes such that the network uplink service coverage probability is maximized. 
Finally, considering millimeter wave channel models, we study the effect of different parameters including minimum data rate requirement, coverage distance and transmit power  on the network performance. As we show, a power allocation schemes with well-tuned parameters can improve the uplink performance of IAB networks considerably. Moreover, with millimeter wave communications and a proper network deployment, the effect of interference on the service coverage probability is negligible.

\end{abstract}

\begin{IEEEkeywords}
5G NR, Integrated access and backhaul, IAB, 3GPP, power control, uplink, service coverage probability, genetic algorithm, TDD, Machine learning, Interference.
\end{IEEEkeywords}

\section{Introduction}

In 5G and beyond, wireless networks will be densified with multiple access points of different types \cite{1},\cite{2}. The access points need to be connected to the operators’ core network via a transport network. On a global scale, fiber and microwave technology are dominant backhauling techniques. Fiber is a reliable link immune to interference and environmental effects with high peak data rate. However, the installation/maintenance cost of fiber may be high and it may not be attainable to deploy it everywhere. Due to the local geometry and features in some locations, fiber installations may not be feasible. Also, there are some governing policies in certain locations that may not allow installing new infrastructures for the use of optical fiber \cite{2}. 

Wireless backhaul, on the other hand, is a scalable and economical backhaul option that can meet the increasing requirements of 5G systems \cite{2}, although it is sensitive to, e.g., blockage, tree foliage, rain, and supports lower peak data rates compared to fiber. For this reason, microwave is a backhaul technology used by most mobile operators worldwide, and the trend is likely to continue in the future. Typical wireless backhaul links are designed for point-to-point communications at 10-80 GHz, with strong line-of-sight (LoS) signal components. Also, even though there are few microwave communication standards, the existing wireless backhaul technologies are mainly based on non-standardized solutions. 

With 5G and beyond, the access links will operate in millimeter wave (mmWave) spectrum, the range which was previously used for backhauling. Thus, there may be a conflict of interest between the access and backhaul links, which requires standardization. On the other hand,  with low-height access points installed on, e.g., lamp posts, there is a probability for blockage, and we also need to support non-line-of-sight (NLoS) communication in the backhaul links. These are the main motivations for the current integrated access and backhaul (IAB) networks \cite{2}. With IAB, the objective is to provide flexible wireless backhauling using 3rd generation partnership project (3GPP) new radio (NR) technology, and provide not only the existing cellular services but also backhaul in the same node and via the same hardware \cite{2},\cite{3}. 

The performance of IAB networks have been studied in different works. In particular,\cite{2} provides the basics for IAB network architecture and studies the service coverage in downlink communication. Then, \cite{3} evaluates the coverage extension improvement in 28 GHz band with IAB deployment in 3GPP urban micro scenarios. Also, \cite{4} investigates the number of IAB nodes that are required for 5G IAB deployment.
Then, \cite{5} investigates the power allocation problem for a proposed in-band self-backhaul scheme using an iterative algorithm. Moreover, \cite{6} investigates power control for moving networks in mmWave based wireless backhaul, and \cite{7} proposes dynamic power control to improve NLOS transmission performance. Additionally, \cite{8} proposes an uplink power control scheme based on machine learning in 5G networks for near-optimum performance in terms of transmit power, data rate, and network energy. In \cite{9}, the IAB networks are studied in an end-to-end manner, in line with 3GPP Release 16 (Rel-16). %proposal and shows that the IAB network has a half-duplex constraint, even with access channel duplexing. 
Interestingly, \cite{10} formulates a multi-hop scheduling problem to offer an efficient IAB network deployment. In \cite{11}, a genetic algorithm (GA) formulation is developed for both IAB node and non-IAB backhaul link distribution. 
%Their work makes a case for efficient routing in locations with severe availability constraints and high blockage densities. 
Finally, \cite{12} investigates the potentials and challenges of mobile IAB, and \cite{13} develops joint scheduling and rate allocation for maximizing the throughput. 

In this paper, we study the effect of power allocation in the uplink performance of IAB networks. Considering mmWave channel characteristics and the power ranges agreed in 3GPP, we develop a GA-based power allocation scheme maximizing the network coverage probability. Here, combined with resource allocation, the UEs and the IAB nodes transmit powers are jointly optimized using GA. Also, we investigate the effect of the interference on the network performance. Moreover, our simulations verify the effect of separate access and backhaul transmission, compared to the cases with simultaneous access and backhaul transmission. Finally, we verify the effect of different parameters such as the minimum data rate requirement, the cell size and the transmit power on the service coverage probability.

Our simulations show that, with a power allocation scheme, the network coverage probability is improved, compared to the cases with non-optimized power allocation. For instance, consider a two-hop IAB network operating at 28 GHz and 400 MHz channel bandwidth. Then, with a service coverage probability of 70\% and typical parameter settings, the implementation of power control leads to a minimum of 5 dB SNR gain, compared to the cases with non-optimized power allocation. Moreover, we have considered and evaluated a case with dedicated slots for backhaul transmission and compared it with a case of having simultaneous access and backhaul transmission. We observed that having dedicated slots for backhaul transmission offers a higher service coverage probability. Finally, for a broad range of parameter settings and mmWave transmission, the effect of interference on the service coverage probability may be negligible, if the network deployment is properly planned.

\section{System Model}

IAB network consists of two types of nodes \cite{2},\cite{4}:
\begin{itemize}
\item IAB donor, consisting of the central unit (CU) and distributed unit (DU) which serves the UEs as well as the other IAB nodes. IAB donor is connected to the core network via a non-IAB, e.g., fiber, backhaul link.
\item IAB nodes, consisting of the DU and mobile termination (MT) units which serves the UEs and, possibly, other IAB nodes in the chain of multi-hop communications. The IAB nodes rely on IAB for backhauling.
\end{itemize}

The motivation for the CU/DU split in IAB donor, as initially suggested in 3GPP Rel. 15 for next generation NodeBs (gNBs), is that time-critical functionalities, such as scheduling, and fast retransmission, can be realized in the DU close to the radio and the antenna, while the less time-critical radio functionalities are centralized in the CU. In Rel. 16 IAB, both out-of-band and in-band backhauling are supported in which the access and backhaul operate in different and the same frequency bands, respectively. In-band backhauling gives the flexibility in resource allocation between the access and backhaul, at the cost of complexity/coordination \cite{14}. In this paper, we concentrate on in-band backhauling.

In an IAB multi-hop chain, the parent node connects to the downstream UEs and IAB nodes via the IAB-DU. The IAB-MT is the module connecting an IAB node to its parent IAB-DU. From many aspects, IAB-MT part of a node behaves like a UE in the sense that it connects to the parent IAB-DU like a UE. %Thus, the parent IAB-DU can control the IAB-MT, in terms of, e.g., transmit power. 
On the other hand, from the UE perspective, the IAB-DU of a node appears as a normal DU. 

\begin{figure}[h]
\vspace{-4mm}
\centering
\includegraphics[width=0.53\textwidth, height= 9cm]{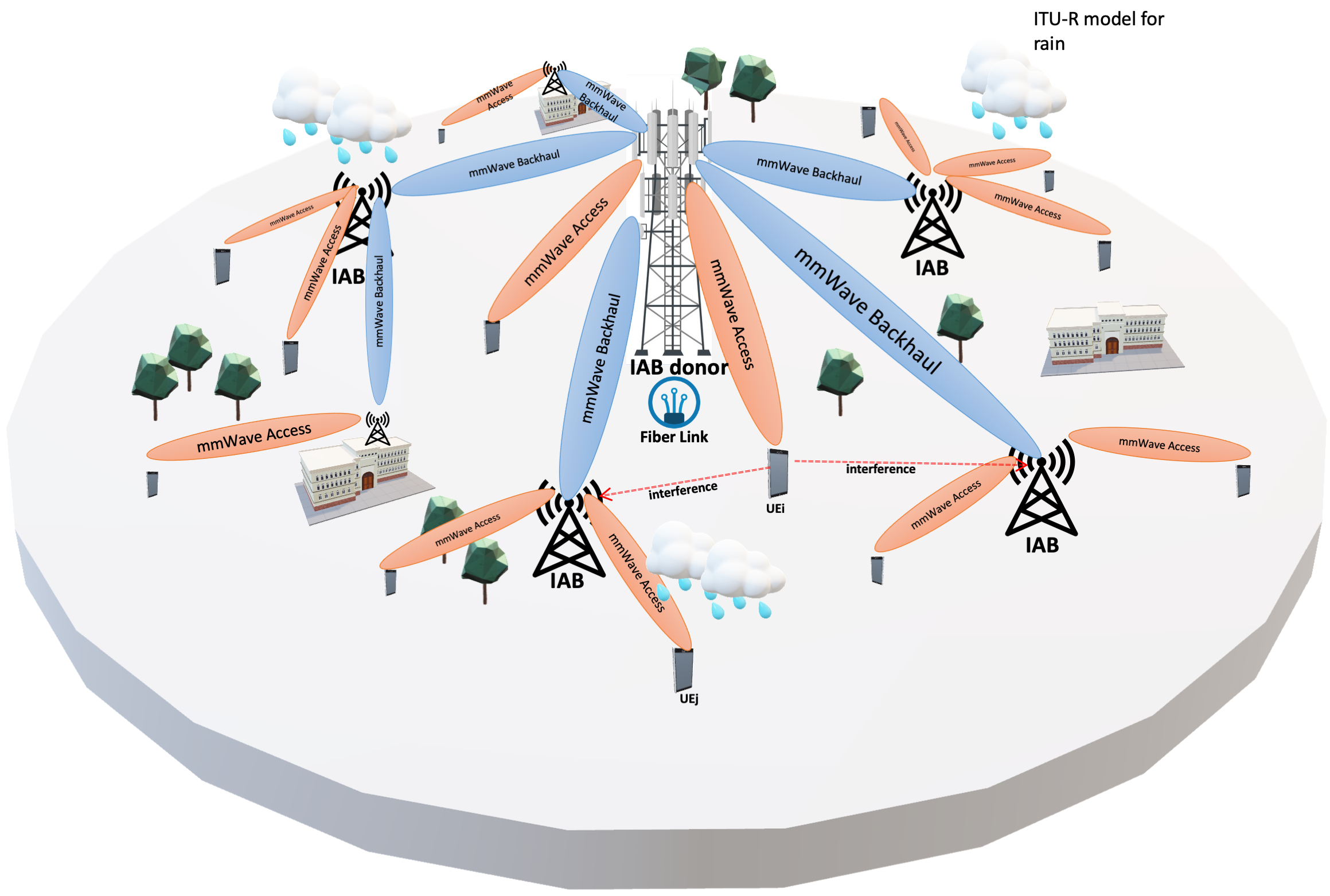}
\vspace{-7mm}
\caption{Schematic diagram of the system model of an IAB network. %There is one IAB donor (MBS) and 4 IAB nodes (SBSs). %The figure also represents the system model for the various simulations carried out.
}
\label{fig.iab_networks}
\vspace{-1mm}
\end{figure}

Figure \ref{fig.iab_networks} shows our considered system model, assuming an outdoor two-hop IAB network. This is motivated by the fact that, as reported by e.g., \cite{2}, \cite{11}, although 3GPP does not limit the possible number of hops, in practice traffic aggregation in the backhaul links and latency become challenging as the number of hops increases.

The system model supports \textit{E} number of UEs that are randomly distributed within the specified coverage area of radius \textit{r}. The UEs are randomly distributed using the finite homogeneous Poisson point process (FHPPP) approach \cite{4}. Also, there are \textit{M} IAB nodes in fixed locations within the coverage area.
Two specific network arrangements are considered: (i) one IAB donor associated with a finite number of stationary IAB nodes, and (ii) two IAB donors associated with a finite number of stationary IAB nodes each, in adjacent macro cells, in order to investigate the effect of inter-cell interference. The wireless channel is modeled to include the effects of shadowing, interference, pathloss, fading and rainfall. 
The following are the general assumptions made in this paper:
\begin{itemize}
\item A central time division duplex (TDD) scheduler governs the communication of the network nodes and UEs.
\item The transmit power of the UEs and IABs is taken as the effective isotropic radiated power (EIRP) as suggested by 3GPP \cite{15}.
\item Periodic access to channel state information (CSI) is available at the IAB donor, IAB node and UE.
\end{itemize}

\subsection{Channel Model}

The received power at either the IAB donor or IAB node is modelled as

\begin{equation}
 P_r =  P_t +  G_t +  G_r -  L -  \sigma -  Y_R - \phi.
\label{eq.3.pr}
\end{equation}

Here, $P_t$ is the transmit power, $G_t$ is the gain of the transmitter, $G_r$ is the gain of the receiver, $L$ is pathloss, $\sigma$ is shadowing loss, $Y_R$ is rain loss, and $\phi$ is the channel fading effect, which is modeled as a Rayleigh flat fading. All values are in dB. In our work, the transmit power values used are EIRP defined in \cite{15}, where $P_\text{EIRP}$ is expressed as

\begin{equation}
 P_{\text{EIRP}} = P_t + G_t.
\label{eq.3.eirp}
\end{equation}

A 3GPP urban macro (UMa) model is selected for the pathloss and shadowing \cite{16}. The model factors in the height of the IAB donor, IAB nodes and the UEs. 
The UMa pathloss model is represented by

\begin{equation}
    \begin{split}
        L = 32.4 + 10 \alpha \log_{10}(d_\text{{3D}}) + 20{\log_{10}{f_c}}\\
        - 10({(d^{'}_{\text{BP}})^2} + (h_{\text{BS}} - h_{\text{UE}})^2).
        \label{eq.3.pathloss}
    \end{split}
\end{equation}

Here, $\alpha$ is the pathloss exponent. 
${d}_{3D}$ is the 3D distance calculated using trigonometric equation, which is the LoS distance from the top of every UE to the top of the base station (BS), i.e, IAB donor or IAB node. 2D distance is the horizontal distance from the BS to a UE. Also, $f_c$ is the carrier frequency, in GHz. The pathloss exponent is related to the signal blockage, either in LoS or NLoS use cases. Also, $h_{\text{BS}}$ is the height of the BS, and  $h_{\text{UE}}$ is the maximum height of the UE, and $d^{'}_{\text{BP}}$ is the break point distance which is determined by the relationship 

\begin{equation}
d^{'}_{\text{BP}} = \frac{4 * h^{'}_{\text{BS}}* h^{'}_{\text{UT}}*{f}_c}{ {c}},
\label{eq.3.breakpoint}
\end{equation}
where, $h^{'}_\text{BS}$ is the effective antenna heights of the BSs, $h^{'}_\text{UT}$ is the effective antenna heights of the UE, and $c$ is the speed of light. 

In mmWave communication, the rain loss may not be negligible, depending on the frequency, distance and rain intensity. The International Telecommunications Union Radio communication (ITU-R) section has a model for corresponding signal attenuation for a given rain rate and operating frequency band \cite{17}. The system model is built on rain rate of between 15 mm/h and 20 mm/h. The ITU-R power-law relationship model is expressed as %Eq. \eqref{eq.3.rain}

\begin{equation}
    Y_R = k * R^{\mathit{\Gamma}}.
    \label{eq.3.rain}
\end{equation}

Here, $Y_R$ is the signal specific attenuation expressed in dB/km, while $R$ is the given rain rate in mm/h. Then, $k$ and $\mathit{\Gamma}$ are polarization coefficients that are determined based on the operating frequency band \cite{15}. Finally, using similar terminology as in \eqref{eq.3.pr}, the interference is expressed as

\begin{equation}
   I_i =  \sum_{\forall j\ne i} P_{\text{EIRP}_j} + G_{r_j} - L_j - \sigma_j - Y_{R_j} - \phi_j.
    \label{eq.3.interferance}
\end{equation}

In this way, using \eqref{eq.3.pr}-\eqref{eq.3.interferance}, the received signal-to-interference-plus-noise ratio (SINR), $\gamma$,  at an IAB node or IAB donor is given by

\begin{equation}
     \text {$\gamma$} = \frac{P_r}{{I}+{ N_0}},
    \label{eq.3.SINR}
\end{equation}
where, $N_0$ is the modeled channel noise. 
Then, the achievable rate, in bits per second, is obtained by

\begin{equation}
    R_b =  \text{BW} * \log_{2}(1 + \text{$\gamma$}),
    \label{eq.3.shanon}
\end{equation}
where BW is the UE channel bandwidth, in Hz, either to the IAB donor or IAB node.  The same concept holds for every IAB node associated with an IAB donor, where one can derive the achievable data rate of the IAB-IAB donor link similar to \eqref{eq.3.shanon}. Moreover, the backhaul data rate of an IAB node is a summation of the data from the UEs that are associated with it, thereby guaranteeing a successful communication of the UEs to the IAB donor.
Then, given a target data rate, $R_b$, the minimum required SINR in a link is found as
\begin{equation}
    {\gamma}_{\text{min}} = 2^\frac{R_b}{\text BW} - 1.
    \label{eq.3.SINR.min}
\end{equation}
Our metric of interest is the service coverage probability. Here, with a two-hop setup, a UE can either connect directly to the IAB donor, or its message is forwarded to the donor IAB via an intermediate IAB node. Then, a transmission fails and a UE is out of coverage if its message can be transferred to the donor IAB in none of these paths, which can be found out via the comparison of the SINR of the links and the minimum required SINR. 

\section{Uplink Power Optimization}
In this section, we formulate the optimization problem regarding the uplink power control in IAB networks. The problem formulation is based on determining appropriate transmit powers for the UEs and associated IAB nodes within a range to meet a pre-defined service requirement determined by a baseline data rate and a corresponding SINR requirement.  

To reduce the optimization complexity, we consider a sequential procedure for resource association and power allocation. Firstly, we start with the UE and IAB node association  rule. Then, we optimize the power allocation based on the considered node association.

Motivated by the 3GPP Rel. 17 discussions on interference management, we evaluate the system performance in two distinct cases where either the access and backhaul transmissions are separated in different time slots, or they can be performed in the same slot. Also, 3GPP considers different ranges of possible transmit powers for the UEs and BSs \cite{15}. 

We apply GA for power control. Here, the GA takes the CSI and the baseline service requirements to implement power allocation for all UEs and IAB nodes, within the specified transmit power range, such that the service requirements of the nodes are satisfied.

In words, the proposed GA follows the following procedure. Let us denote the number of iterations by $N$, determined by the designer.
First, $K$ possible random solutions are selected. This set is created by containing randomly selected transmit powers, with each value coming from the pre-defined range of transmit power values supported by the UE and IAB node. For each set of the $K$ possible solutions, the SINR value of the links and the network coverage probability are calculated. The best solution, leading to maximum coverage probability, which we refer to as Queen in the following, is determined and kept for the next generation. Keeping the Queen for the next generation guarantees the continuous improvement in different iterations of the GA.

In the next step, $S$ $<<$ $K$  sets of transmit power values are generated around the Queen, by making  few mutations on it, i.e., changing few elements inside the Queen, such that the UEs/IABs transmit powers remain within their acceptable range. Then, $V=K-S-1$ random set of solutions are generated. The algorithm iterates $N$ times and the final Queen is returned as the best solution maximising the coverage probability.

As an advantage, the proposed algorithm is generic in the sense that it can be applied in different channel models and UEs/IABs transmit power ranges. Moreover, as we show in the following, the proposed GA converges with a few iterations. This is important because, it is straightforward to show that the considered power allocation is an NP-hard problem with no closed-form or easy-to-search solution. As opposed, the proposed GA only requires $NK$ solution checkings to reach a (semi) optimal solution.

\vspace{-1mm}
\begin{algorithm}
\caption{GA-Transmit Power Control Algorithm}\label{alg:euclid}

For each instance, with a set of UEs and IAB nodes, do the followings:
\begin{algorithmic}[1]
\STATE Create $K$, e.g., $K=10$, sets of transmit powers, randomly selected from the range between the minimum and maximum transmit powers allowed for the UEs and the IAB nodes. Each element in the set corresponds to either a UE or an IAB node transmit power. 
\STATE For each set, calculate the received power, SINR and coverage probability.
\STATE Find the set with the highest coverage probability, and call it as the Queen. The Queen is kept for the next generation.

\STATE Create $S{\ll}K$, e.g., $S=5$, sets of transmit powers around the Queen. These sets are created by small modifications in the Queen such that the UEs/IABs transmit powers remain within their possible ranges. 
\STATE Create $V = K - S -1$  sets of random transmit powers, with each element in the set related to the allowed transmit power range for the UE and the IAB nodes. 
\STATE Go to back to   \textbf {Step} \textbf 2 and continue for $N$ iterations.
\STATE Finally, return the Queen as the optimal power solution which has highest coverage probability after $N$ iterations.
\end{algorithmic}
\end{algorithm}

\section{Simulation Results}
The model is deployed for an urban macro environment with 3GPP and ITU-R parameters detailed in Table \ref{Parameters} \footnote{The parameter settings are not necessarily aligned with the Ericsson parameters of interest.}.
The simulation results are presented in different parts as follows.
\begin{table}[h!]
\vspace{-2mm}
\centering
\small\addtolength{\tabcolsep}{-4pt}
\caption{Simulation Parameters.}
\label{Parameters}
%\resizebox{\textwidth}{!}{%
\begin{tabular}{|l|r|l|r|}
\hline
\textbf{Parameter}    & \textbf{Value}              & \textbf{Parameter}            & \textbf{Value}          \\ \hline
$f_c$      & 28 GHz                       & Cell radius ($r$)       & 200 m          \\ \hline
$BW$              & 400 MHz                     & Shadowing ($\sigma$)              & 4 dB           \\ \hline
Sub-carrier spacing  & 120 kHz                     & Pathloss Exp.($\alpha$)     & 4              \\ \hline
Min. RBs                 & 24                         & Eff. Ant. Height      & 1 m            \\ \hline
Max. RBs                 & 270                          & Ref. distance ($d_0$)    & 1 m            \\ \hline
Thermal noise ($T_{0}$)    & -174 + BW (dB) & Num. IAB nodes         & 4 per cell \\ \hline
${\text{UE}}_{\text{Noise figure}}$  ($N_f$)         & 5 dB                        & Min. Data Rate    & 64 kbps \cite{18}       \\ \hline
Noise ($N_0$)     & $T_0$ + $N_f$ (dB)        & UE EIRP         & \{23 - 43\} dBm  \\ \hline
${\text{Height}}_{\text{IAB Donor}}$ & 25 m                        & IAB node EIRP        & \{35 - 53\} dBm  \\ \hline
${\text{Height}}_{\text{IAB Node}}$ & \{21 - 24\} m                 & Receiver gain & 25 dB\cite{18}          \\ \hline
UE height   & 1.5 m                       & Rain rate ($R$)     & \{15 - 20\} mm/h \\ \hline
\end{tabular}
\end{table}

\begin{figure}[h]
\vspace{-1mm}
\centering
\includegraphics[width=0.5\textwidth, height= 6cm]{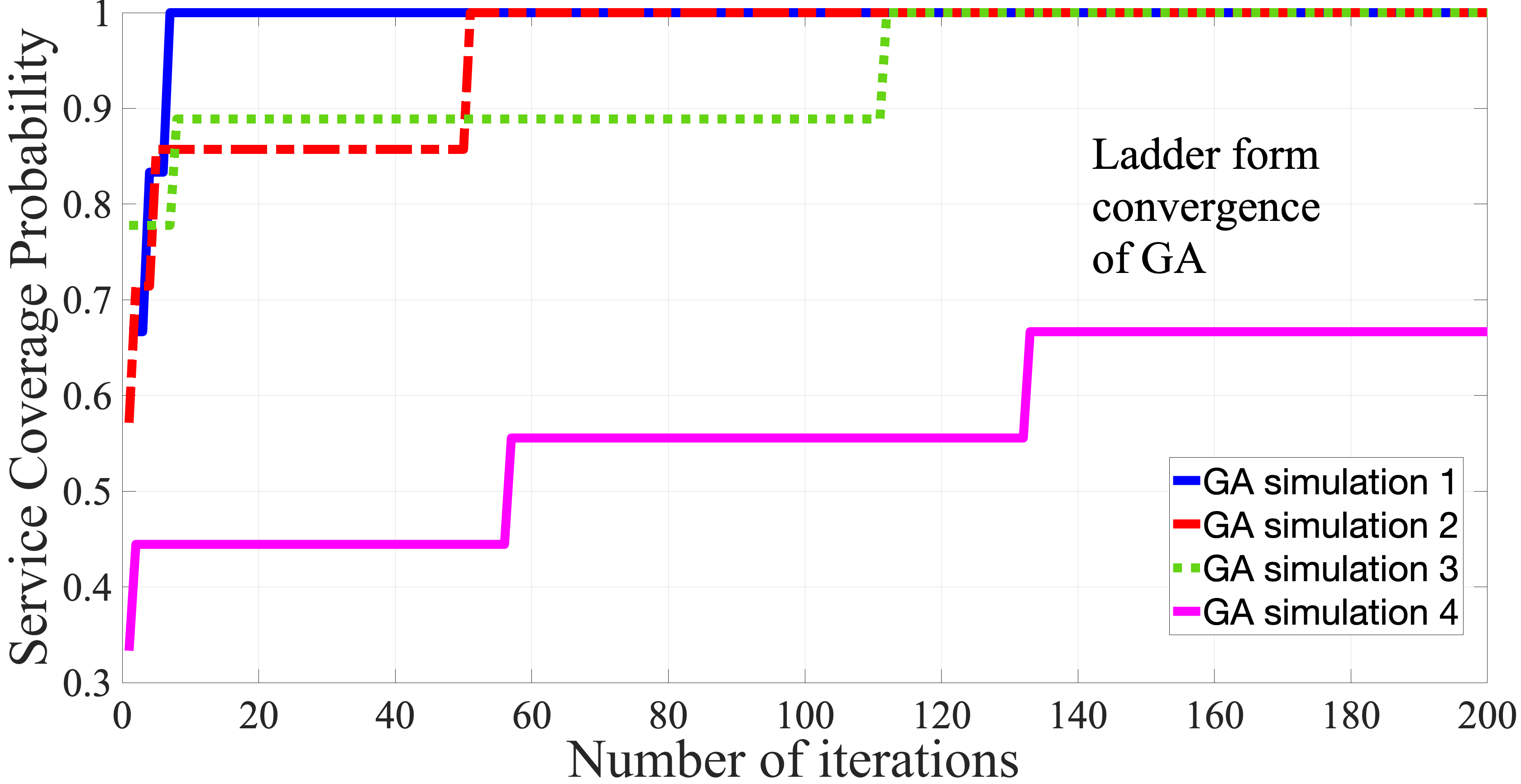}
\vspace{-7mm}
\caption{Examples of the convergence of the GA.}
\label{fig.gaconverge}
\vspace{-2mm}
\end{figure}

Figure \ref{fig.gaconverge} shows different examples of the convergence of the GA. 
Here, we set $N$ = 200, $K$ = 20 and $S$ = 10. As seen in Fig. \ref{fig.gaconverge}, the algorithm converges in a ladder form. The observed ladder form is because the GA may not necessarily find the best solution in every iteration, and it may be trapped for a while in a local minimum. However, due to Step 5 of the Algorithm, GA can always avoid a local minimal and reach the global optimum if sufficiently large number of iterations is considered \cite{19}, \cite{20}. In GA, the larger the number of iterations, the more accurate the final solutions but at a cost of the running time. Then, as seen in Fig. \ref{fig.gaconverge} and our various non-included simulations, in different channel realizations the proposed GA converges with a few iterations. This reduces the optimal complexity, compared to, e.g., exhaustive search based solutions, significantly.

\begin{figure}[h]
\vspace{-3mm}
\centering
\includegraphics[width=0.5\textwidth, height=6cm]{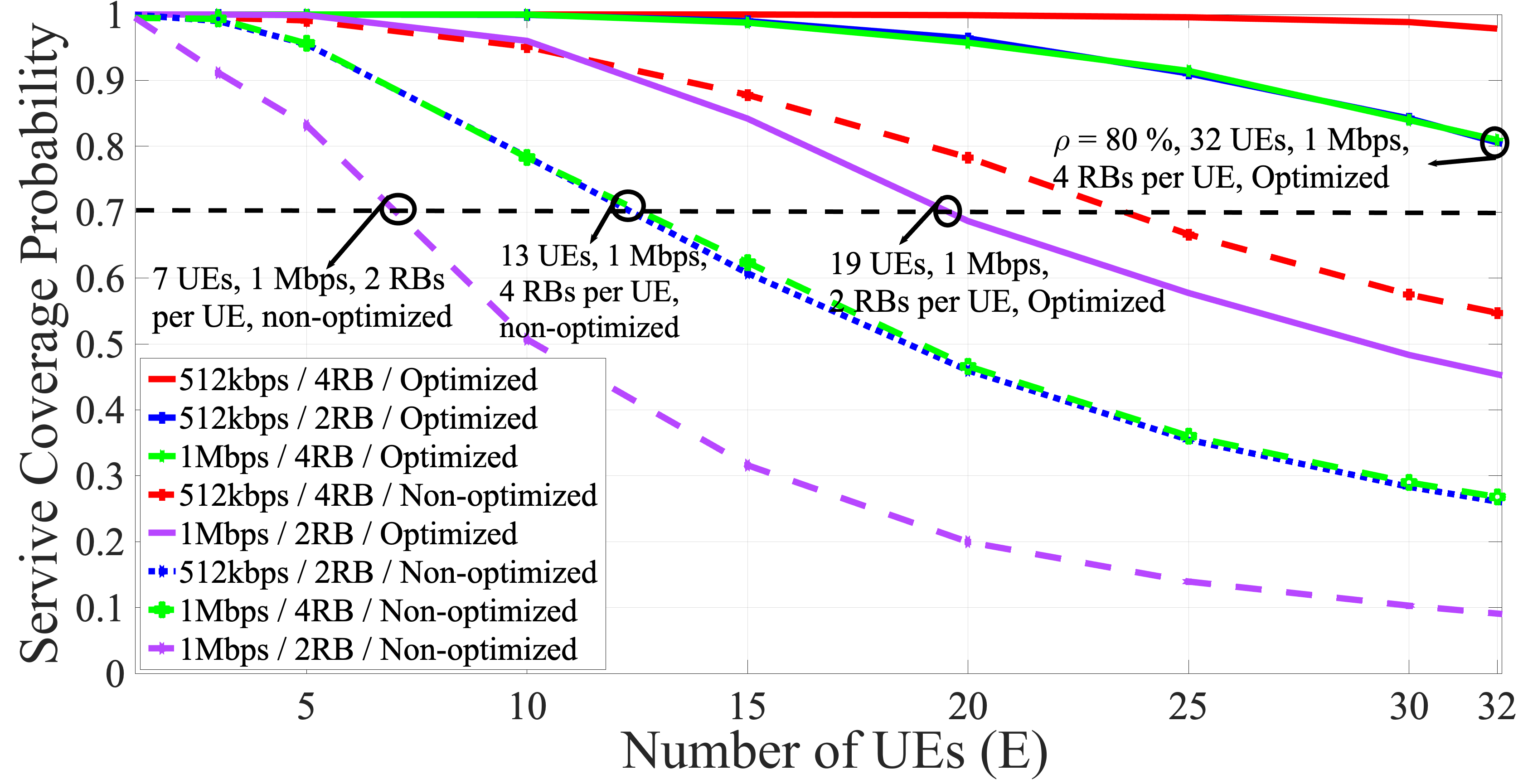}
\vspace{-7mm}
\caption{Service coverage probability versus the number of UEs with 2 and 4 RBs per UE.}
\label{fig.SC2COMB}
\vspace{-2mm}
\end{figure}

In Fig. \ref{fig.SC2COMB}, we investigate the number of UEs that can be served within the cell, with and without uplink power optimization. We also investigate the effect of increasing the resource blocks (RBs) per UE when the access and backhaul links are operating at different times. Here, we consider two adjacent cells, with a fixed cell radius.

From Fig. \ref{fig.SC2COMB}, it is observed that an optimal power allocation makes it possible to serve considerably higher number of UEs. For instance, consider the parameter settings of Fig. \ref{fig.SC2COMB} and service coverage probability of $70\%$. Then, with a minimum data rate 1 Mbps and 2 RBs per UE, the number of supported UEs increases from 7 with non-optimized power allocation to 19 UEs in the cases with optimal power control. This is indeed at the cost of dynamic coordination at the IAB donor, which may not be possible in practice. The relative gain of power control increases with the number of RBs per UE. Also, when the number of concurrent UEs within the cell increases, the service coverage probability decreases. 
This may be attributed to the increase in the interference within the cell since the number of transmitters, i.e., the UEs, increases.

\begin{figure}[h]
\vspace{-3mm}
\centering
\includegraphics[width=0.5\textwidth, height=6cm]{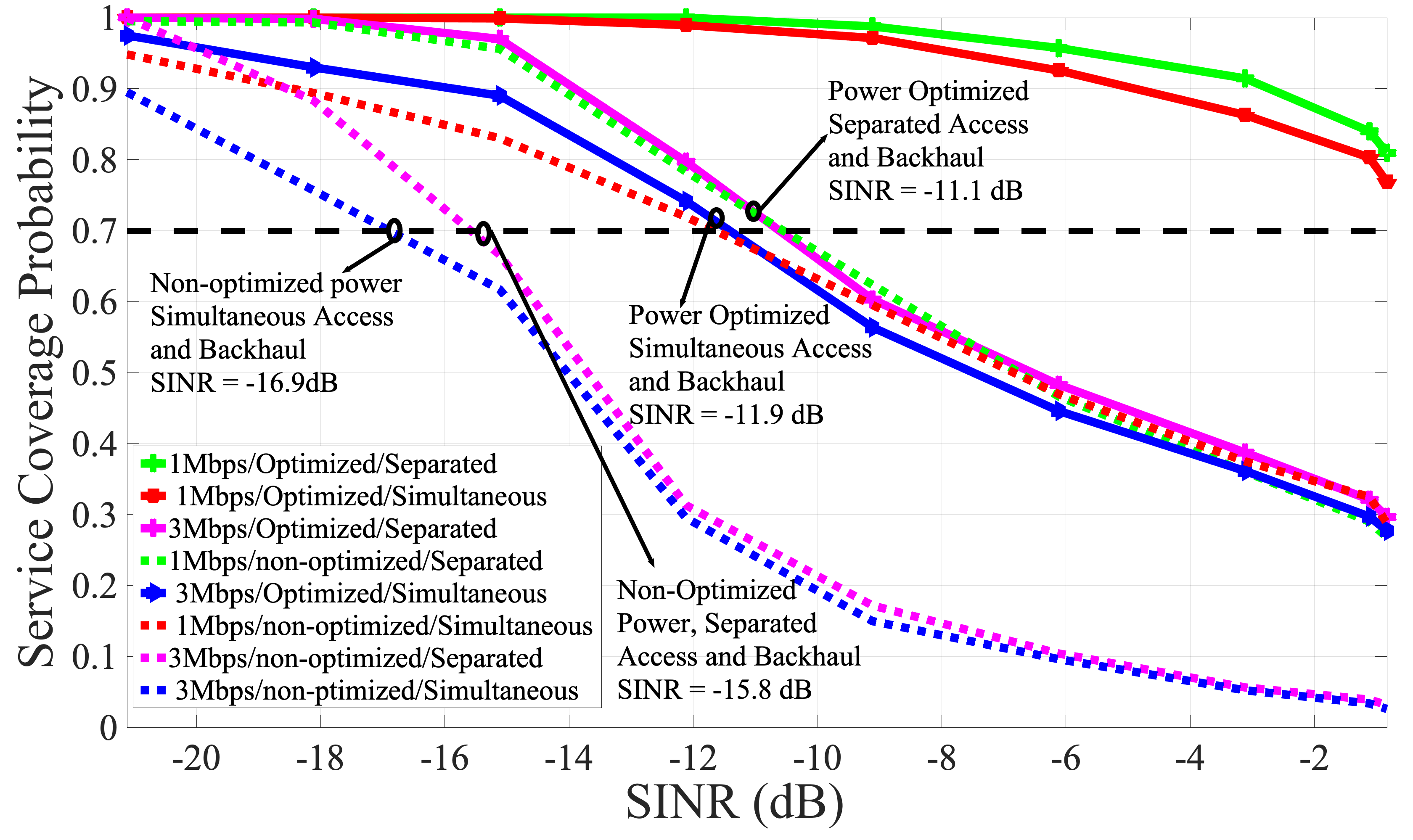}
\vspace{-7mm}
\caption{Service coverage probability versus the SINR with 4 RBs per UE.}
\label{fig.sinrdifftdd}
\vspace{-2mm}
\end{figure}

In Fig. \ref{fig.sinrdifftdd}, we study the system performance in the cases with simultaneous and separated operation of the access and backhaul links in the uplink slots. Here, the results are presented for both cases with non-optimized and optimized power allocation.

Figure \ref{fig.sinrdifftdd} shows that, when access and backhaul are separated, with service coverage probability $70\%$, there is a minimum of 5 dB gain in SINR after the implementation of power optimization, compared to non-optimized power allocation. 
Furthermore, when the access and backhaul links are working simultaneously, there is a decrease in the SINR.
With the parameter setting of Fig. \ref{fig.sinrdifftdd} and service coverage probability 70\%, the simultaneous operation of the access and backhaul in the uplink slots reduces the SINR, compared to the cases with separated access and backhaul operation, by 1 dB. Here, it should be noted that although separating the access and backhaul in the uplink slots saves the uplink signals of the IAB-connected UEs from the high transmit power of the IAB-MTs, still the neighbour non-IAB networks will suffer from the IAB-MTs high transmit powers.
\begin{figure}[h]
\vspace{-3mm}
\centering
\includegraphics[width=0.5\textwidth,height=6cm]{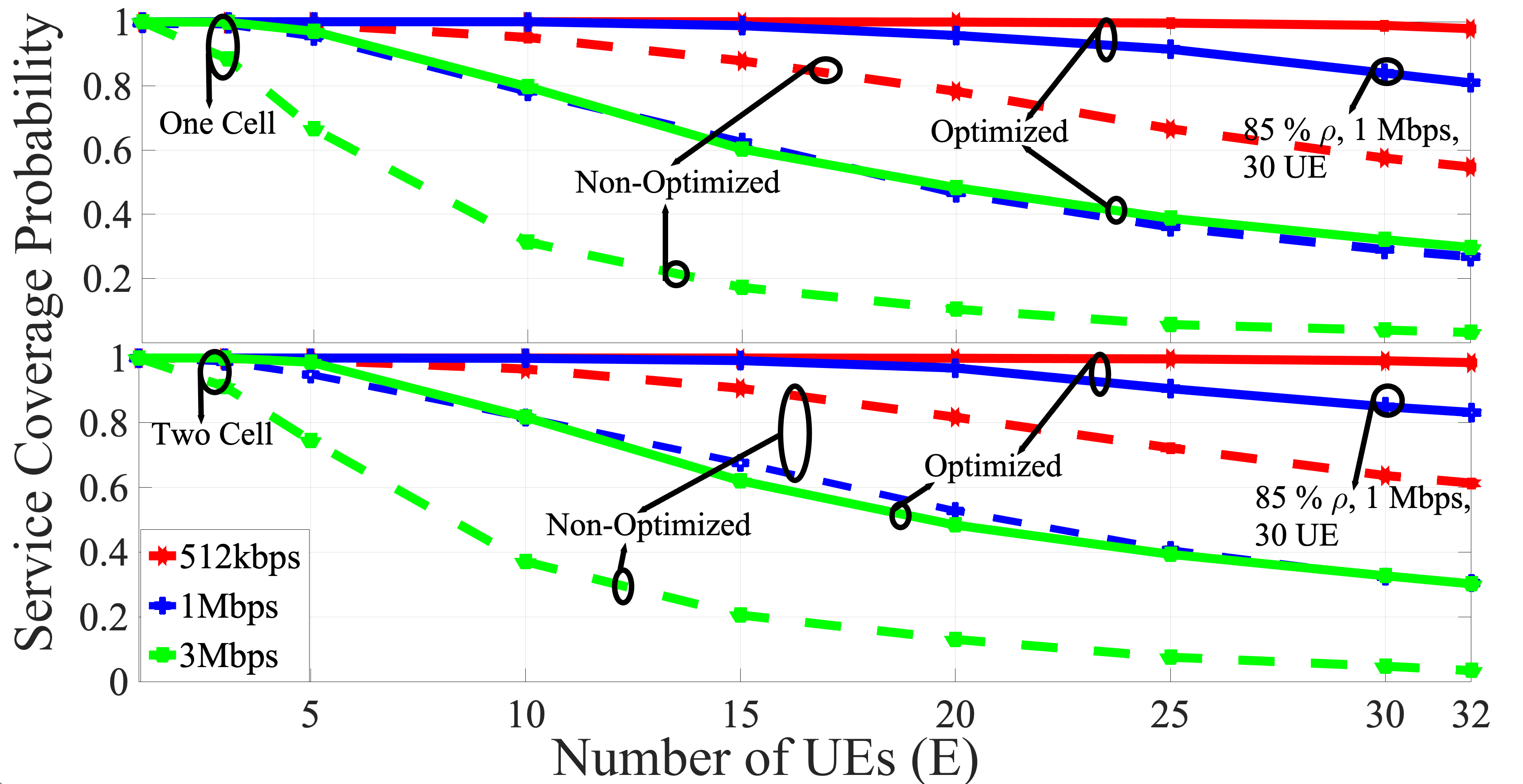}
\vspace{-7mm}
\caption{Inter-cell interference. This figure shows a comparison between service coverage with one cell in comparison to the cases with two-cell set-up.}
\label{fig.intercell}
\vspace{-2mm}
\end{figure}

Figure \ref{fig.intercell} shows the effect of inter-cell interference on the number of UEs that can be served in the cell, with and without uplink power optimization. The figure compares service coverage probability of a single-cell set-up with two adjacent cell set-up. There is not much difference in the results from the two considered scenarios. Notably with a sufficient cell radius, coming from, e.g., network planning, the effect of inter-cell interference is minimal. This is because at 28 GHz, the signal strength decreases rapidly with the distance, whereby the effect of inter-cell interference is minimal if the distance between the cells is sufficiently large.

\begin{figure}[h]
\vspace{-3mm}
\centering
\includegraphics[width=0.5\textwidth, height=5.4cm]{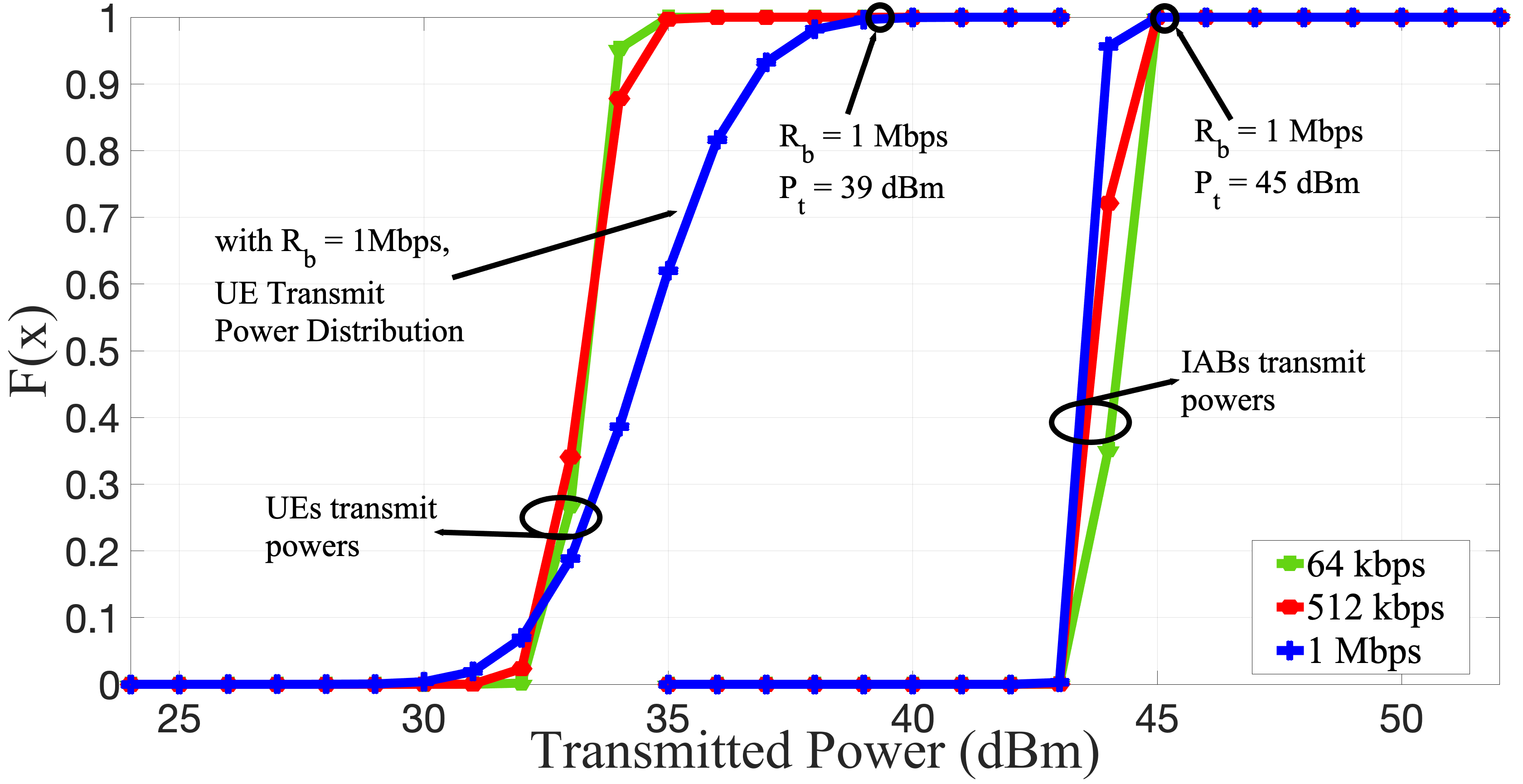}
\vspace{-7mm}
\caption{CDF for UEs and IABs transmit powers.}
\label{fig.cdfsbs}
\vspace{-2mm}
\end{figure}

Figure \ref{fig.cdfsbs} studies the cumulative distribution function (CDF) of the UEs and the IAB nodes transmit powers, for different data rates. The results are presented for the cases with the UEs and IAB nodes transmit powers optimized via the GA. Also, the transmit powers are optimized within the range specified by 3GPP \cite{15}.
According to the figure, the required transmit power increases slightly with the data rate. However, there is relatively the same transmit power distribution of all the UEs irrespective of their associated BS, either IAB donor or IAB nodes. 
Also, the transmit power variation of the UEs increases as the data rate increases. . 
On the other hand, the transmit power variation of the IABs is low for the considered set of UEs data rates, as the IABs use almost the same transmit power. That is, 
compared to the IABs, the UEs transmit power is more sensitive to the data rate.

\section{Conclusion}

We studied the effect of uplink power control on the service coverage probability of two-hop IAB networks. We developed a GA-based scheme for power control, and studied the effect of different parameters on the network performance. 

Our results show that compared to the cases with non-optimized power allocation, our proposed sequential node association and power control algorithm improves the energy efficiency and the service coverage probability of IAB networks. Also, for different network configurations, the proposed GA converges with a few iterations, which reduces the optimization complexity compared to, e.g., exhaustive search based schemes, significantly.
Moreover, separating the IAB and UEs transmissions in the uplink slots may give the chance to improve the service coverage probability, as the IAB donor-connected UEs signals are not affected by the IABs high transmission powers. Finally, with proper network planning and mmWave communications, the effect of interference on the network performance is not considerable.

\section{Acknowledgement}

This work was supported in part by VINNOVA (Swedish Government Agency for Innovation Systems) within the VINN Excellence Center ChaseOn and in part by the European Commission through the H2020 project Hexa-X (Grant Agreement no. 101015956).

%\begin{thebibliography}{00}

\vspace{12pt}

\end{document}